\documentstyle[prl,twocolumn,aps,epsf]{revtex}

\begin{document}
 \tolerance 50000

\draft

\twocolumn[\hsize\textwidth\columnwidth\hsize\csname@twocolumnfalse\endcsname

\title{A Density Matrix Renormalization Group Approach to \\
an Asymptotically  Free Model with Bound States}

\author{M.A. Mart\'{\i}n-Delgado$^{1}$ and  G. Sierra$^{2}$,  
} 
\address{
$^{1}$Departamento de F\'{\i}sica Te\'orica, Universidad Complutense ,
Madrid, Spain.  
\\ 
$^{2}$Instituto de Matem{\'a}ticas y F{\'\i}sica Fundamental, C.S.I.C.,
Madrid, Spain. 
}

\maketitle 

\begin{abstract} 
We apply the DMRG method to the 2 dimensional delta function
potential which is a simple quantum mechanical model with
asymptotic freedom and formation of bound states.
The system block and the environment block of the DMRG  
contain the low energy and high energy degrees 
of freedom, respectively. The ground state energy and the 
lowest excited states
are obtained with very high  accuracy. We compare the DMRG method with
the Similarity RG method and propose its generalization
to field theoretical models in high energy physics. 
\begin{center}
\parbox{14cm}{}

\end{center}
\end{abstract}

\vspace{-0.8 true cm}

\pacs{
\hspace{2.5cm} 
}

\vskip2pc]
\narrowtext

\vspace{-1 true cm}

A hallmark of an asymptotically free theory such as QCD 
is that it contains many degrees of freedom,  with very 
different energy scales, which are  
coupled by the interaction Hamiltonian. 
Perturbative methods
are valid for short distance physics but they
fail for small momentum transfers or for energy
scales where the bound states are formed. 
The existence of multiple energy scales suggests that
the Renormalization Group approach is the 
correct strategy  to 
attack  these non perturbative problems. 
In recent years there
has been several proposals to 
extract effective low energy Hamiltonians 
using RG methods. 
Of particular interest is 
the light-front Hamiltonian approach
advocated in references \cite{wilson,perry}
which uses a similarity RG method (SRG) \cite{gw1,wegner}. 
In this method the 
RG flow is given by an unitary transformation which diagonalizes the
Hamiltonian by succesive elimination of 
 the off-diagonal matrix elements.
The SRG-cutoff can be seen as the width
of the band which contains 
the non vanishing  off-diagonal matrix elements of the
Hamiltonian. At the end of the SRG flow the width is zero
and the corresponding Hamiltonian contains in its diagonal
all the eigenvalues of the original one.

In this letter we shall propose an alternative RG approach
to study asymptotically free models using the 
Density Matrix Renormalization Group (DMRG). We shall
also show the relations and differences between 
the DMRG and the
SRG methods. The DMRG was proposed by White in 1992 to solve  
the problems of the old  real space RG methods encountered
in the 70's, which led in those days 
to their abandon 
in favour of Montecarlo techniques \cite{white1}.   
The DMRG has by now become a 
standard numerical RG method applied
to many body problems in Condensed Matter and 
other branches of Physics ( see references 
\cite{white2,nw} for reviews).
It is thus  challenging
to test  how the DMRG  handles the subtle
dynamics of asymptotically free theories. 
To our knowledge this is the first paper devoted to the
subject. For this reason we have 
choosen as a theoretical lab   a simple model
possesing  the essential properties 
of asymptotic freedom and formation of bound states,
which are shared by realistic theories like QCD.

The natural candidate for such a simple model is 
provided by a 2d quantum mechanical particle 
subject to a delta function potential \cite{jackiw}. 
The  solution of the 2d delta function 
Schr\"odinger equation 
requires a regularization and renormalization schemes  as
in an  ordinary quantum field theory. We shall
use for our purposes the lattice regularization
introduced by Glazek and Wilson in their SRG study of the
problem \cite{gw2,gw3}. 
These authors formulated the problem
in momentum space where the states are labelled by 
an integer  $n$ that ranges between an infrared cutoff
$M$ and an ultraviolet cutoff $N$ (i.e. $M \leq n \leq N$). 
The kinetic energy $E_n$ of the state $n$ increases
exponentially as $E_n = b^{2 n}$, where $b$ is an arbitrary
constant greater than one. For numerical computations
we shall take the value $b= \sqrt{2}$ as in references 
\cite{gw2,gw3}.
The interaction Hamiltonian between the states $n$ and $m$ 
is given by $- g \sqrt{E_n E_m}$, where $g$ is the coupling 
constant of the problem. The discrete lattice Hamiltonian $H$
is defined by the matrix elements

\begin{equation}
H_{n m} = \delta_{n,m} b^{2 n} - g \; b^{n+m}, \;\; M \leq n , m \leq N
\label{1}
\end{equation}

\noindent
An overall   shift of the levels by a constant term, i.e.
$n \rightarrow n + n_0$, implies that $H_{n m}$  
scales with the factor $b^{2 n_0}$. This is
a discrete version of scale invariance, which is  broken by 
the infrared and ultraviolets cutoffs $M$ and $N$.  
The latter symmetry implies that all the scales
contribute to the observables, which makes very hard
an accurate determination of their value  by 
methods other than the exact one. 




The first step in the DMRG method is the partition of
the system
in two pieces called the system block
and the environment block \cite{white1}. 
In our case we shall choose the system block $B^L_{\ell}$
to be given by  the low energy levels $n$ which lie between
the infrared cutoff $M$ and the scale $\ell$ (i.e. 
$M \leq n \leq  \ell$),
while the environment  block $B^H_\ell$ will contain
the high energy levels $n$ between the ultraviolet cutoff $N$ and the
scale $\ell$ (i.e. $\ell \leq  n \leq N$). The whole system,
with energy levels ranging  from $M$ to $N$, is obtained
as the ``superblock'' $B^L_\ell \bullet \circ B^H_{\ell+3}$, where
$\bullet $ and $\circ$ are the $n=\ell+1$ and $n=\ell +2$ energy levels
respectively (see fig.1).

\begin{figure}
\hspace{-0.8cm}
\epsfxsize=9cm \epsffile{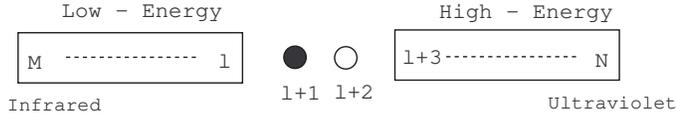}
\narrowtext
\caption[]{ Superblock decomposition of the energy scales.
}
\label{fig1} 
\end{figure}
\noindent

The  parameter $\ell$ varies from $M$
to  $N-3$ and it labels the  DMRG flow.
Let us supose we want to find the
GS of the whole system.  
We shall choose a trial GS wave function 
$\psi_\ell(n)$ as follows,

\begin{equation}
\psi_\ell (n) = \left\{  \begin{array}{lccl} 
a_1 L_\ell(n) & & &M \leq n \leq \ell \\
a_2  & & & n= \ell+1 \\
a_3 & & & n= \ell +2 \\
a_4 R_{\ell +3}(n) & & & \ell+3 \leq n \leq N \end{array} \right.
\label{3}
\end{equation}

\noindent where $L_\ell$ ( resp. $R_{\ell+3}$) is a 
normalized  vector
which describes the contribution of the low ( resp. high) energy
block $B^L_\ell$ ( resp. $B^H_{\ell +3}$) to the GS of the 
superblock  $B^L_\ell \bullet \circ B^H_{\ell+3}$. 
The ansatz (\ref{3}) is the momentum space version
of the real space DMRG applied by White to study
a free particle in a box \cite{white2,nw}. 
Our approach is close in spirit to the momentum space
DMRG method proposed by Xiang \cite{xiang}.  
The energy of the state (\ref{3}) can be conveniently written as

\begin{equation}
\langle \psi_\ell |H| \psi_\ell \rangle = 
\langle {\bf a}| H_{SB}(\ell) | {\bf a} \rangle
\label{4}
\end{equation}

\noindent   where $|{\bf a}\rangle$ is the vector 
$(a_1,a_2, a_3,a_4)$  
and the superblock Hamiltonian $H_{SB}(\ell)$
is the $4 \times 4$ matrix given by

\begin{equation}
H_{SB}(\ell) = \left( \begin{array}{cccc}
h_L & h_{L \bullet} & h_{L \circ} & h_{L H} \\
h_{L \bullet} & h_\bullet & h_{\bullet \circ} & h_{H \bullet} \\
h_{L \circ} & h_{\bullet \circ} & h_\circ & h_{H \circ} \\
h_{L H} & h_{H \bullet} & h_{H \circ} & h_H \end{array} \right) 
\label{5}
\end{equation}

\noindent whose entries read

\begin{equation}
\begin{array}{ll}
h_L =\langle L_\ell | H | L_\ell \rangle, & 
h_H = \langle R_{\ell+3} | H | R_{\ell +3} \rangle\\
h_\bullet = H_{\ell+1, \ell+1}, & h_\circ =
h_{\ell +2, \ell+2} \\
h_{L \bullet} = \sum_{n=M}^\ell H_{n,\ell+1} L_\ell(n), & 
h_{L \circ} = \sum_{n=M}^\ell H_{n,\ell+2} L_\ell(n) \\
h_{H \bullet} = \sum_{n=\ell+3}^N H_{n,\ell+1} R_{\ell +3}(n),&
h_{L H} = \langle L_{\ell} | H| R_{\ell+3} \rangle \\
h_{H \circ} = \sum_{n=\ell+3}^N H_{n,\ell+2} R_{\ell +3}(n) &
h_{\bullet \circ} = H_{\ell+2,\ell+3}  
\end{array}
\label{6}
\end{equation}

\noindent  where $H_{n, m}$ are the matrix elements 
given in eq. (\ref{1}). Notice that
eq.(\ref{4}) takes the form of an eigenvalue problem
in a reduced vector space with only 4 degrees of freedom. The GS
of the superblock can be found by looking for the
lowest eigenvalue $E_{1}(\ell)$ 
 of the $4\times 4$ matrix  $H_{SB}$. 
The variational nature of the  construction
gives an upper bound of the exact GS energy.
If the vectors $L_\ell$ and $R_{\ell +3}$ 
coincide with the low energy and high energy pieces
of the exact GS wave function then the   
DMRG algorithm presented so far
would reproduce the exact result. 
Of course  this is not in general the case but 
nevertheless, one can actually use the DMRG algorithm to 
improve in succesive steps the GS energy. 
The idea is to apply a continuity argument.
Suppose we shift the scale $\ell$ to the next  
high energy level, say $\ell +1$. Then the new low 
energy vector $L'_{\ell +1}$ will be related to the previous
one $L_\ell$ by the equation

\begin{equation}
L'_{\ell+1}(n) = \left\{ \begin{array}{llll}
                a'_1 L_{\ell}(n) & & & M \leq n \leq \ell \\
                a'_2 & & & n = \ell+1 \end{array} \right.
\label{8}
\end{equation}

\noindent  where $(a'_1,a'_2)  = (a_1,a_2)/ \sqrt{a_1^2 + a^2_2}$ 
is the normalized two component vector
obtained by the projection of the lowest eigenvalue of
$H_{SB}(\ell)$ into the block $B^L_{\ell} \bullet$. 
Similarly  the energy $h'_L(\ell+1)$ associated to the
latter block is given by

\begin{equation}
h'_L(\ell+1) = ( a'_1, a'_2) \left( \begin{array}{ll}
h_L(\ell) & h_{L \bullet}(\ell) \\
h_{L \bullet}(\ell) & h_\bullet (\ell) \end{array} \right) 
\left( \begin{array}{c} a'_1 \\ a'_2 \end{array} \right)
\label{9}
\end{equation}

The data $L'_{\ell +1}$ and $h'_L(\ell+1)$ fully 
characterize  the new  block $B'^L_{\ell+1}$   
which can be regarded  as the renormalization of 
the block 
$B^L_{\ell} \bullet$. The next step is to construct 
the superblock $B^L_{\ell +1} \bullet \circ B^H_{\ell +4}$
which by the same techniques leads to the construction of 
a new block $B'^L_{\ell+2}$ and so on. This procedure
is iterated until the scale $\ell = N-3$, 
where one reverses the DMRG steps in order to update
the high energy blocks $B^H_\ell$ using the low energy blocks
built in the previous steps. After a few sweeps from low to high
energy and viceversa the lowest eigenvalue of the superblock 
Hamiltonian (\ref{5}) converges to a fix value which gives
the DMRG estimation of the GS energy.  To start out
the process one has to grow up the system to its actual
size. This can be done by considering  superblocks of the form
$B^L_{M+p} \bullet \circ B^H_{N-p}$ where $p=0, \dots, 
(N-M-3)/2$. The last value of $p$ yields a system containing  
all the  scales from $M$ to $N$. 
The low and high energy blocks
constructed in the warm up are the starting 
point for  the sweeping procedure explained 
above (see \cite{white2,nw} for details). 
The previous algorithm has been  generalized 
in reference \cite{msn} to find out not only the GS but 
the low lying excitations as well.

Let us now present our DMRG results for the case
considered in references \cite{gw2,gw3}, where $M= -21, N=16$
and $g= 0.06060600032108866$. The latter value
of $g$ is choosen in such a way 
that the exact ground state energy of (\ref{1}) 
is given exactly by $-1$.  
 The DMRG algorithm presented above gives 
the exact ground state energy 
with an error of $10^{-14}$ (see table 1). 
Using the extension of the DMRG proposed
in \cite{msn} we have also computed 
the GS and the lowest 
3 excited states
of the Hamiltonian (\ref{1}). In table 1 
we compare our DMRG results with the 
exact ones in terms of the relative
deviation

\begin{equation}
\delta E_n = \frac{ E_n (DMRG) - E_n (exact)}{ E_n (exact)}
\label{error}
\end{equation}

\begin{center}
\begin{tabular}{|c|c|c|c|c|} 
\hline
$n$   & 1   & 2 & 3 & 4  \\ 
\hline
$\delta E_n$  & $7 \times 10^{-15}$ & 
$1.04 \times 10^{-7}$
& $3.36 \times 10^{-6} $ & 
$ 1.41 \times 10^{-6}$ \\ 
\hline
\end{tabular}
\end{center}
\begin{center}
Table 1. 
Relative error $\delta E_n$ of the four lowest eigenstates
of the Hamiltonian (\ref{1}).  
\end{center}

As shown in table 1 the accuracy of the 
excited states energies is lower than that 
of the GS. This feature is peculiar to the
delta function Hamiltonian and it does 
not arise for the quantum mechanical models
studied in \cite{msn}. 
There are several  reasons for the very high 
accuracy of the DMRG
applied to the Hamiltonian (\ref{1}):
i) the DMRG 
gives a variational upper bound to the exact GS energy which 
is usually improved in every DMRG step;
ii) all the
matrix elements of the whole Hamiltonian are used many times
to feedback  the superblock so 
that no information is lost;
iii) the DMRG method focus on the determination of the
GS and the low lying states.

In fig.2 we plot the DMRG wave function
which after the third sweep 
is indistinguishable
from the exact one. 

\begin{figure}
\hspace{-0.8cm}
\epsfxsize=8cm \epsffile{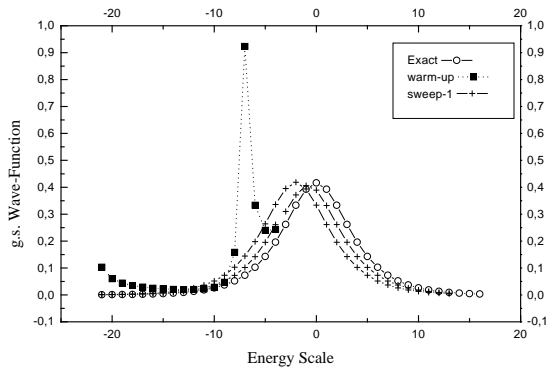}
\narrowtext
\caption[]{DMRG v.s. exact wave function.
}
\label{fig2} 
\end{figure}
\noindent

It is interesting to investigate the nature of the DMRG flow
as compared with the one of the similarity RG method. 
In the SRG the effective Hamiltonian $H(s)$ 
evolves as a function of $s$ according to the
Wegner equation \cite{wegner},

\begin{equation}
\frac{d H(s)}{d s}  = [ [H_d(s), H(s)], H(s)]
\label{10}
\end{equation}

\noindent where $H_d(s)$ is the diagonal part of $H(s)$. 
The initial condition of eq.(\ref{10}) is $H(0)=H$, where
$H$ is the original Hamiltonian of the problem. 
The parameter $s$ ranges from $0$ to $\infty$ and it 
can be identified with the inverse 
square of the energy width $\lambda$, i.e. $s= 1/\lambda^2$.
 Eq. (\ref{10}) implies that $H(s)$ is related to $H$ 
by an unitary transformation and  therefore they share their
eigenvalues. When $s$ increases, the off diagonal matrix elements
of $H(s)$, located at distances  greater
than the width   $\lambda=1/\sqrt{s}$, become very small.
When $s=\infty$ the effective Hamiltonian $H(\infty)$
is diagonal and  all its entries coincide 
with the eigenvalues of $H$. 
The numerical integration of eq.(\ref{10}) 
requires of course to follow the evolution of 
all the entries of
$H$.  One would like instead  to project the effective
Hamiltonians to smaller (``window'')
matrices in order to reproduce the bound state eigenvalue \cite{gw3}. 
In a sense, the superblock $4\times4$ matrices (\ref{5})
resemble the window matrices of ref. \cite{gw3}.
Motivated by the SRG ideas \cite{gw3}, 
we have studied the RG flow of the 
eigenvalues $E_i(\ell) (i=1,\dots,4)$ 
of the superblock Hamiltonian  (\ref{5}).
In fig.3 we plot the lowest eigenvalue $E_1$,
together with the remaining ones scaled down
by a factor $b^{2 \ell}$. We can clearly see 
from fig.3 
that $E_1$ stays constant through all the DMRG steps
while $E_i (i=2,3,4)$ vary  
with the energy scale $b^{2\ell}$ with some deviations
depending on the energy region. The plateaus correspond
to low energy regions while the oscillations and bumps
occur for  intermediate and high energies. 
To a first order approximation,  which is almost exact
for the plateaus, the superblock Hamiltonian 
(\ref{5}) can be written as

\begin{equation}
H_{SB}(\ell) = O_\ell \left( \begin{array}{cccc}
E_1 & 0 & 0 & 0\\
0 & E'_2 b^{2 \ell} & 0 & 0 \\
0 & 0 & E'_3 b^{2 \ell} & 0 \\
0 & 0 & 0 & E'_4 b^{2 \ell} \end{array}
\right) O^\dagger_\ell
\label{11}
\end{equation}

\noindent where $O_\ell$ is a unitary matrix. 
Using eq.(\ref{11}) one can show that the superblock
Hamiltonians satisfy the following second order 
recursion relation,

\begin{eqnarray}
& H_{SB}(\ell) = & \label{12} \\ 
& \frac{1}{b + b^{-1}}
\left( b^{-1} U_\ell H_{SB}(\ell+1) U^\dagger_\ell +
b U^\dagger_{\ell-1} H_{SB}(\ell-1) U_{\ell -1} \right) & 
\nonumber
\end{eqnarray}

\noindent where $U_\ell = O_\ell O^\dagger_{\ell +1}$. 
The continuum limit of eq.(\ref{12}) gives the flow equations

\begin{equation}
H_1 \equiv \frac{d H_{SB}}{d \ell} - [ \eta, H_{SB}], \; \;
\frac{d H_1}{d \ell}  = [ \eta,  H_1] 
\label{13} 
\end{equation}

\noindent where $\eta= \frac{ d O_\ell}{d \ell} O^\dagger_\ell$.
Eq.(\ref{13}) is a second order diferential equation
which is to be  
compared with the first order equation eq.(\ref{10}).
The DMRG flow is  a sort of similarity transformation
with some eigenvalues running with the scale. 
Using  the standard RG terminology 
the lowest eigenvalue $E_1$ can be associated
with a marginal operator while the eigenvalues
$E_i$ for $i=2,3,4$ are associated with infrared irrelevant 
operators which vanish at the fixed point Hamiltonian
$H_{SB}(\ell = M)$. Indeed all the entries of 
$H_{SB}(\ell = M)$ are very small except for the entry
$h_{H}= - 0.999$ whose value is close to the
bound state energy. These results suggest that
the exactness of the DMRG method is due to a careful
treatment of the irrelevants operators, which 
in other RG methods are difficult 
to control in general.

From a conceptual point
of view the DMRG offers a new way of thinking 
about cutoffs and RG flows in high energy physics.
Traditional cutoffs remove high energy states 
while the lowering of the cutoff produces 
effective operators for lower energies \cite{wilson65}. 
In the Lagrangian formulation this strategy
can be implemented perturbatively without
much difficulty. However in the Hamiltonian
formulation it gives rise to small denominators
problems involving energy differences between
the states kept and the states truncated in the
RG process \cite{bh,gw2}. This latter problems do not arise
in the DMRG truncation for it uses a non perturbative
self-consistent algorithm to find the best choice
of the effective Hilbert spaces and Hamiltonians.

The next step in the application of the DMRG to high
energy physics is of course to consider  
field theoretical models with asymptotic freedom
and bound states. 

\begin{figure}
\hspace{-0.8cm}
\epsfxsize=8cm \epsffile{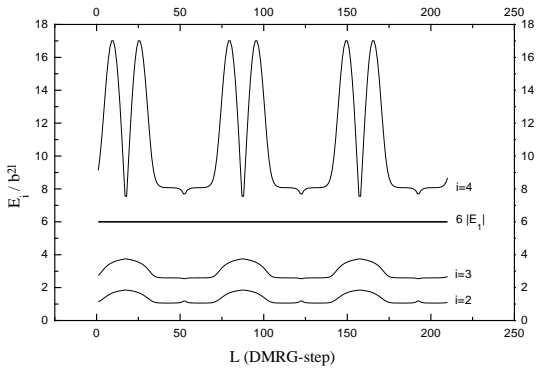}
\narrowtext
\caption[]{Plot of the rescaled  superblock eigenvalues
$E_i/b^{2\ell}$ ($i=2,3,4$) as a function of the DMRG step L
from sweep 2 to 4. 
}
\label{fig3} 
\end{figure}
\noindent

The most appropriate formalism for this application of the
DMRG is known as Discrete Light-Cone Quantization in momentum space
(DLCQ) \cite{dlcq}, \cite{schwinger}. In the DLCQ approach
the Hilbert space is finite dimensional and 
the light-front 
Hamiltonian $H_{LC}$ acting on it 
is similar to that of a many-body Hamiltonian
in condensed matter \cite{schwinger}. The search for bound
states amounts to solving the Schroedinger equation

\begin{equation}
H_{LC} |\psi \rangle = M^2 \; |\psi \rangle
\label{13}
\end{equation}

\noindent where $M^2$ is the mass of the bound state. 
 Thus,  one can apply to (\ref{13})
standard diagonalization techniques such as Lanczos. 
The DMRG method allows us
to study larger Hilbert spaces than those achieved with 
the Lanczos method.
This is needed in order to recover the continuum 
limit of $H_{LC}$.

The key to make the clever truncation of
states in the RG process is given by the density matrix
of the blocks. This is a systematic procedure with no
wild guessing about the wave function. A given block
will contain the most 
representative multiparticle and momentum  states
in order to reconstruct the lowest lying  bound states. 
On the other hand, the DMRG is a  numerical method, unlike the more
analytical SRG method \cite{glazek}, and 
does not need perturbative
inputs. 
The basic  requirement for the  DMRG method to work 
 is a discretized
Hamiltonian acting on finite dimensional Hilbert spaces. 
As a  first test of this program we have solved eq.(\ref{13})
for the Positronium state of the massive and massless
Schwinger model in the one-fermion sector with the DMRG. 
These results will be reported elsewhere.

Thefore, we expect that 
the main ideas presented in this letter
can be generalized to the DLCQ Hamiltonians.  
Specifically, 
the breaking of the system into low
energy and high energy blocks which are constantly 
updated through the DMRG process. 
On the other hand
the DMRG method combined  with the DLCQ  approach does 
not have the sign problems that 
emerge in  the Montecarlo methods used in 
Lattice Gauge Theories \cite{schwinger}.
In summary, we believe that the power  shown in condensed matter 
systems by the DMRG method is worthwhile to be translated into Particle Physics.

{\bf Acknowledgements} 
This work was supported by the DGES spanish grant
PB97-1190. 

\vspace{-0.5 true cm}


\begin{thebibliography}{99}
\vspace{-1.5 true cm}

\bibitem{wilson} K.G. Wilson et al.
, Phys. Rev. D 49,
6720 (1994).

\bibitem{perry} R.J. Perry, nucl-th/9901080. 

\bibitem{gw1} St. D. Glazek and K.G. Wilson, Phys. Rev. D 48,
5863 (1993); 49, 4214 (1994).

\bibitem{wegner} F. Wegner, Ann. Physik 3, 77 (1994).



\bibitem{white1} S.R. White, Phys. Rev. Lett. 69, 2863 (1992),
Phys. Rev. B 48, 10345 (1993).

\bibitem{white2} S.R. White,  Phys. Rep. 301, 187, (1998).

\bibitem{nw} R. Noack and S.R. White, 
Lecture Notes in Physics, I.Peschel, X. Wand, K. Hallberg, eds.
(Springer-Verlag, 1999).


\bibitem{jackiw} R. Jackiw, in M.A.B. Beg Memorial Volume,
A. Ali and P. Hoddbhoy, eds. (World Scientific, Singapore, 1991).


\bibitem{gw2}  K.G. Wilson and St. D. Glazek, in the Procs.
of the Ninth Physics Summer School at the Australian National
University, H.J. Gradner and C.M. Savage, eds. (World
Scientific, Singapore, 1997). 

\bibitem{gw3} St. D. Glazek and K.G. Wilson, hep-th/9707028.



\bibitem{xiang} T. Xiang, Phys. Rev. B 56, 5061 (1996).


\bibitem{msn} M.A. Mart\'{\i}n-Delgado, G. Sierra and R. Noack,
cond-mat/9903100. 

\bibitem{wilson65}K. G. Wilson, Phys. Rev. {\bf 140}, B445 (1965).

\bibitem{bh} C. Bloch and J. Horowitz, Nucl. Phys. 8, 91 (1958).

\bibitem{dlcq}
H.-C. Pauli and S. J. Brodsky, Phys. Rev. D {\bf 32}, 1993, 2001
(1985).

\bibitem{schwinger}
 T. Eller, H.-C.Pauli, and S. J. Brodsky, Phys. Rev. D {\bf 35}, 1493 (1987). 
For a review, see 
  S. Brodsky, H-C Pauli, S. Pinsky, Phys.Rept. 301 (1998) 299-486.

\bibitem{glazek} St. D. Glazek, Acta Phys. Pol. {\bf B29} (1998) 1979-2064,
hep-th/9712188.




\end{thebibliography}
\end{document}